\documentstyle[12pt]{article}
\begin{document}
\begin{titlepage}
\hfill{PURD-TH-92-12}
\vskip 2cm
\centerline{\Large \bf TOP CONFINEMENT}
\vskip 1cm
\centerline{Bijan Haeri, Jr.}
\vskip 1cm
\centerline{\it Department of Physics}
\centerline{\it Purdue University}
\centerline{\it West Lafayette, Indiana 47907-1396}
\vskip 1.5cm
\centerline{ABSTRACT}
 \noindent
A new interaction that confines the top quark at the electroweak scale is
proposed to account for a top quark mass
potentially close in value to the electroweak scale.
\end{titlepage}

 The standard model predicts the existence of the top quark,
but fails to describe why the top mass $m_{top}>100$ GeV, could be comparable
 to the electroweak scale $O(100\ GeV)$.
Here we explore the possibility of presently unknown
interactions at the electroweak scale
giving mass to the top quark, and forming
a top condensate that contributes to a dynamical breaking of
electroweak symmetry.
 To accomplish this, we propose top condensation via a confining mechanism.

An example of quark condensation in nature is approximate
$SU(2)_{L}\times SU(2)_{R}$ dynamical chiral symmetry breaking (DCSB) in
quantum chomodynamics (QCD),
 where
quark condensate formation leads to chiral symmetry breaking and the
production of nearly massless, approximate Nambu-Goldstone particles (pions).
QCD confinement is a sufficient mechanism for DCSB, and in the process
of producing pions, the quark constituents of
the pions have their masses dynamically generated with a value on order
of the confinement scale.

Motivated by DCSB in QCD, we introduce a confining mechanism
(in addition to QCD) for the top quark.
This top-confining interaction
is an $SU(3)_{T}$ gauge symmetry, which confines the top quark
near the
electroweak scale, thus dynamically generating a top mass, $m_{t}$, on
the order of the electroweak scale.

The third family of quarks and leptons have the following transformation
properties under $SU(3)_{T}\times SU(3)_{C}\times SU(2)_{W}$ (where
$SU(3)_{C}$,
and $SU(2)_{W}$ are the QCD color, and weak gauge symmetries respectively):
\begin{eqnarray}
&[(t, b)_{L}\rightarrow (3,3,2)], \nonumber \\
              & [t_{R}\rightarrow (3,3,1)], \nonumber \\
                     &[b_{R}\rightarrow (1,3,1)], \nonumber \\
           &[(\tau, \nu _{\tau})_{L}\rightarrow (1,1,2)], \nonumber \\
            &[\tau_{R}\rightarrow (3,1,1)].
\end{eqnarray}
The above assignment is free of the $[SU(3)_{T}]^{2}U(1)_{H}$ trace over
hypercharge anomaly
(where $U(1)_{H}$ is the weak hypercharge gauge symmetry), but
does contribute to the $[SU(3)_{C}]^{2}U(1)_{H}$, and $[SU(2)_{W}]^{2}U(1)_{H}$
 trace over hypercharge
anomalies, the $[U(1)_{H}]^{3}$ anomaly, and the $[SU(3)_{C}]^{3}$ and
$[SU(3)_{T}]^{3}$ anomalies, requiring
the introduction of right-handed $SU(2)_{W}$ singlet, and  left-handed
$SU(2)_{W}$ doublet
particles transforming under yet another new interaction (besides the
top-confining one), $SU(2)_{A}$, which is also confining at a scale greater
than the top-confining scale (for reasons described below).
These additional particles transform under $SU(3)_{T}\times
SU(2)_{A}\times SU(3)_{C}\times SU(2)_{W}\times U(1)_{H}$
as:
\begin{eqnarray}
                  &[q_{R}\rightarrow (1,2,3,1,-{2\over{3}})], \nonumber \\
                  &[l_{L}\rightarrow (1,2,1,2,-1)], \nonumber \\
                  &[p_{R}\rightarrow (3,2,1,1,0)].
\end{eqnarray}

Since $SU(2)$ representations are real, we can replace one $SU(2)_{A}$
doublet (out of two) of $l_{L}$ fermions with their conjugate fields, and the
result
is equivalent to having a right handed $SU(2)_{A}$ doublet. The $l_{L}$
fermions
are then part of a chiral symmetric strongly interacting vector theory, namely
one flavor technicolor (it has been shown that strongly
interacting  chiral
$SU(2)$ theories with an even number of flavors
like this one have their chiral symmetry dynamically broken
\cite{hsu}). We take advantage of this fact to
give the $SU(2)_{A}$ interaction a large enough scale $O(1\ TeV)$
to (along with $SU(3)_{T}$) dynamically break the electroweak symmetry.

The top-confining interaction insures the presence of $SU(3)_{T}$ singlet
boundstates below the electroweak scale. But it is still necessary to
introduce 2 sets of scalar particles that transform under $SU(3)_{T}\times
 SU(3)_{C}\times SU(2)_{W}$ as:
\begin{eqnarray}
                  &[\phi_{1}\rightarrow (3,1,1)], \nonumber \\
                  &[\phi_{2}\rightarrow (1,1,2)],
\end{eqnarray}
to reproduce the proper low energy phenomenology for the $\tau$, $\nu_{\tau}$,
and b-quark particles (and the other 2 families of quarks and leptons).
 The $\phi_{1}$ scalars will confine with
$b_{L}$ and $\tau_{R}$ to form
$SU(3)_{T}$ singlet composite massless particles; we assume that chiral
$SU(3)_{T}$ interactions preserve global
 currents which involve chiral fermions with no opposite chirality
 partners, and therefore
$b_{L}$ and $\tau_{R}$, do not become dynamically massive \cite{af}.
The resulting composite chiral particles can form boundstates
with their respective fundamental $SU(3)_{T}$ singlet
right and left-handed partners, namely $b_{R}$ and $\tau_{L}$
 through Yukawa couplings. The top quark has both left and right-handed
chiral partners under $SU(3)_{T}$ and participates in vector interactions,
and thus becomes dynamically massive. The top quark also forms an $SU(3)_{T}$
singlet massive boundstate with the $\phi_{1}$ scalars, with a mass
approximately equal to the $SU(3)_{T}$
 confinement scale, $\Lambda$; this composite fermion is
the top quark of the standard
model.
As long as the vacuum expectation value of $\phi_{1}$,
 $<\phi_{1}>$,  is much smaller than $\Lambda$,  there will be
no spontaneous symmetry breaking of the $SU(3)_{T}$ gauge symmetry.
 The $\phi_{2}$ scalars are necessary to form the Yukawa sector for the
3 families of quarks and leptons (except the top quark, which becomes massive
through the $SU(3)_{T}$ dynamics).
We are free to choose $<\phi_{2}>$ to be small relative to the electroweak
scale, because in this model the $\phi_{2}$ field plays no role
 in breaking the
$SU(2)_{W}$ gauge symmetry (as in the standard model).

The top-confining force is strong at the electroweak scale, therefore
we can approximately ignore
 weak hypercharge, weak gauge, and color interactions, giving
an $SU(6)_{L}\times SU(4)_{R}\times U(1)_{V}$ global
 symmetry for
$(t_{L}, b_{L})$ and $(t_{R},\tau_{R})$.
Dynamical symmetry breaking takes place when the top-confining dynamics
generate mass
for the $t$ quark.
leaving behind an $SU(3)_{L}\times SU(3)_{V}\times U(1)_{R} \times
U(1)_{V}$ global symmetry.
The $SU(3)_{L}\times U(1)_{R}$ part of the global symmetry corresponds to
composite massless fermions that satisfy the 't Hooft anomaly
matching conditions \cite{thooft}, namely the $b_{L}, \phi_{1}$ composite
particle
has the same
 $[SU(3)_{L}]^{3}$ anomaly contribution as $b_{L}$, and similarly the
$\tau_{R}, \phi_{1}$ composite particle anomaly contribution to
 $[U(1)_{R}]^3$
 is
equal to that of $\tau_{R}$.
The spontaneous symmetry breaking generates
35 Nambu-Goldstone particles, 3 of which are pion-like and
make small contributions to breaking electroweak symmetry by becoming
part of the longitudinal
components of the $W^{+}$, $W^{-}$, and $Z$ particles.
 The remaining 32 particles are
pseudo-Nambu-Goldstone particles made up of superpostions
color octets, leptoquark
color triplets, and an electric charge neutral axion \cite{fs}.
$q_{R}$ and $l_{L}$ similarly have an approximate
 $SU(2)_{L}\times SU(6)_{R}\times U(1)_{V}$
 global symmetry,
that is broken (at a much higher scale, O(1\ Tev), than $\Lambda$) to give an
additional $SU(2)_{W}$ triplet of pion-like Nambu-Goldstone particles,
which also contribute to the electroweak symmetry breaking, and another
35 pseudo-Nambu-Goldstone particles.

At this point we must draw on methods developed for confining vector
gauge
theories (ie. QCD) to calculate the top quark mass and the top-pion
form factor $f_{t-\pi}$.
This top-confining model is a hybrid between a vector gauge theory
and a chiral gauge theory. Under the $SU(3)_{T}$ transformations
$t_{L}$ has a right-handed partner $t_{R}$, and therefore  is associated
with a vector current interaction, while $b_{L}$ ($\tau_{R}$) has no right-
(left-) handed
partner under $SU(3)_{T}$, and participates in a chiral current.
As mentioned previously, we assume the chiral symmetries of $b_{L}$ and
$\tau_{R}$ are preserved at the top-confining scale.
To evaluate the phenomenological restrictions on this model, we need to
calculate the top mass, and the $\rho$ parameter. The approximations
used to derive the
S parameter of
Peskin and Takeuchi \cite{pt}
are not applicable, since $\Lambda$ is too small,
of order of the electroweak scale, rather than the $O(1\ TeV)$ necessary
for a calculation of S.
There are numerous methods of modeling confinement in QCD, which give
expressions for the dynamical quark mass function \cite{munem}. Rather than
use a complicated model, we choose a
very simple model which captures the essential features of the
dynamical quark mass function: vanishing in the ultraviolet, and of the
order of the confinement scale in the infrared. We obtain similar
values for the dynamical quark mass, and the pion decay constant (using
the equation of Ref.\ 7) of QCD
as the other models \cite{haeri}.
For the top-confining interaction this model is given by (in Euclidean space):
\begin{eqnarray}
           M_{t}(p^{2}) & =\Lambda,\quad -p^{2}<\Lambda^2 \nonumber \\
                        & =0,\quad -p^{2}>\Lambda^2,
\end{eqnarray}
\begin{eqnarray}
       M_{b_{L}}(p^{2}) & =0,\ \ 0<-p^{2}<\infty,
\end{eqnarray}
where $m_{t}\equiv M_{t}(0)=\Lambda$ is the dynamical top quark mass,
while no mass is generated for the $b_{L}$ quark mass.
 The difference between the $t$ and $b_{L}$
quark masses stems from the $t$ and $b$ quark isospin
symmetry breaking due to the top-confining
interaction, and leads to a non-zero contribution to the $\rho$ parameter,
given by,
\begin{equation}
\delta\rho_{T}={{{f_{t-\pi}^{\pm}}^{2}-{f_{t-\pi}^{0}}^{2}}
\over{F_{0}^{2}}},
\label{ceq}
\end{equation}
where $\delta\rho_{T}$ is the contribution from $SU(3)_{T}$ $t$, and $b$
isospin symmetry breaking to $\rho-1\equiv \delta\rho$,
 and $f_{t-\pi}^{\pm}$, and $f_{t-\pi}^{0}$ are the charged, and neutral
top-pion decay constants respectively. $F_{0}$ is the total neutral decay
constant,
$F_{0}=\sqrt{{F_{l-\pi}^{0}}^{2}+{f_{t-\pi}^{0}}^{2}}\simeq 246 GeV$, where
$F_{l-\pi}^{0}$
is the A-pion (formed from $SU(2)_{A}$ $l_{L}$ fermions) decay constant,
and $F_{l-\pi}^{0}=F_{l-\pi}^{\pm}$ as a result of $t$ and $b$ isospin
 symmetry being respected by the $SU(2)_{A}$ interactions.
 About
$96\%$ of $F_{0}$ has to come from $F_{l-\pi}^{0}$ with
its corresponding scale $O(1\ TeV)$,
 because the $SU(3)_{T}$ scale of $\Lambda$ is too small for the
top-confining interactions to give
a significant contribution to $F_{0}$.
Using the ladder approximation and Eq.\ 5 to calculate
 $f_{t-\pi}^{\pm}$ and $f_{t-\pi}^{0}$ \cite{atew}, Eq.\ 6  becomes
(in Euclidean space) \cite{tt},
\begin{equation}
\delta\rho_{T}={9\over{16 \pi^{2} F_{0}^{2}}}\int^{\infty}_{0} dk^2
 {M_{t}^4(k^2)\over{(k^2+M_{t}^{2}(k^2))^{2}}},
\label{feq}
\end{equation}
giving
\begin{equation}
\delta\rho_{T}={9\over{32\pi^{2}}}\Lambda^2.
\label{qeq}
\end{equation}
Eq.\ 7 is not reliable in the infrared regime.
We take the actual value of the contribution to
$\delta\rho_{T}$
be within $50\%$ of the value found from Eq.\ 7 (the error in $\delta\rho_{T}$
is difficult to estimate and may be larger; we calculate
$\delta\rho_{T}$ as a rough guide to the size of the $t$ and $b$ quark
isospin symmetry breaking due to the $SU(3)_{T}$ interactions).
The isospin splitting also leads to a contribution to $\delta\rho$ from
the mass splitting in the charged color octet pseudo Nambu-Goldstone
bosons (PNGB)
 given by \cite{at}:
\begin{equation}
\delta\rho_{PNGB}=\epsilon\left(
{1\over{2}}\int^{1}_{0} dy\Delta_{t} \ln({{\Lambda^{2}}\over{\Delta_{t}}}) +
{1\over{2}}\int^{1}_{0} dy\Delta_{b_{L}}
 \ln({{\Lambda^{2}}\over{\Delta_{b_{L}}}})
 - M_{P\pm}^{2}\ln({{\Lambda^{2}}\over{M_{P\pm}^{2}}})\right),
\label{eeq}
\end{equation}
where
\begin{eqnarray}
    \Delta_{t} &= M_{Pt}^{2}+(1-y)(M_{P\pm}^{2}-M_{Pt}^{2})\nonumber\\
    \Delta_{b_{L}} &=
M_{Pb_{L}}^{2}+(1-y)(M_{P\pm}^{2}-M_{Pb_{L}}^{2})\nonumber
\end{eqnarray}
\begin{equation}
\epsilon=3{{g^{2}+g\prime ^{2}}\over{32\pi^{2}M_{Z}^{2}}},
\end{equation}
and $M_{P\pm}$, $M_{Pt}$, and $M_{Pb_{L}}$ are the charged, neutral composed of
$t$, and neutral composed of $b_{L}$ colored octet PNGB's.
We take the experimental lower bound for the color octet PNGB's to be
approximately $100\ GeV$, and expect the masses to be no larger than about
 ${2\over{3}}\Lambda$. For values of $\Lambda$ between $150\ GeV$ and
$180\ GeV$, we find $\delta\rho_{PNGB}$ to be small comppared to
 $\delta\rho_{T}$.

For $\Lambda=150\ GeV$, we have $m_{t}=150\ GeV$, and
 $.5\%<\delta\rho_{T}<1.5\%$. The contribution to
 $\delta\rho$ from standard model radiative
corrections, $\delta\rho_{SM}$, is about $0.7\%$, while experimentally
the largest
possible values of $\delta\rho=\delta\rho_{SM}+\delta\rho_{T}$
are approximately $1.3\%$ and $1.6\%$ at $90\%$ and $68\%$ confidence level
respectively\cite{ros}; making the
lower range of $\delta\rho_{T}$
phenomenologically acceptable. These numbers for $\delta\rho_{T}$ are
not to be taken too seriously, except to indicate that the $SU(3)_{T}$ breaking
of $t$ and $b$ isospin symmetry may be small enough to be in agreement with
experiment.

By introducing an additional confining interaction for the top quark,
$b_{L}$, and $\tau_{R}$ of the  third family of quarks and leptons, and
a pure one flavor technicolor sector,
we have found a value of top mass comparable to the electroweak scale, and
dynamically broken the electroweak gauge symmetry.
In addition to $O(100\ GeV)$ pseudo-Nambu-Goldstone particles, there will be
massive boundstates involving $t$, $b_{L}$, and $\tau_{R}$
at around  $\Lambda$.
The origin of the current masses for the known
5 flavors of quarks is not addressed.
\vskip 1cm

\centerline{\bf Acknowledgements}
\noindent
The author is grateful for contributions to this work by M. B. Haeri.
The author also benefited from discussions with T. Takeuchi,
S.T. Love, H. Lew, J. M. Cornwall, and
C.D. Roberts.
This work was supported by the US Department of Energy.

\end{document}